\documentclass[10pt, journal, twocolumn]{IEEEtran}

\ifCLASSINFOpdf
\else
\fi

\usepackage{amssymb}
\usepackage{bm}
\usepackage{graphicx}
\usepackage{diagbox}
\usepackage{multirow}
\usepackage{threeparttable}
\usepackage{cite}
\usepackage{amsmath,amsfonts}
\usepackage{algorithmic}
\usepackage{algorithm}
\usepackage{array}
\usepackage{color}
\usepackage[caption=false,font=normalsize,labelfont=sf,textfont=sf]{subfig}
\usepackage{textcomp}
\usepackage{stfloats}
\usepackage{url}
\usepackage{upgreek}
\usepackage{verbatim}

\begin{document}
	%
	\title{\Large Joint Sparsity Pattern Learning Based Channel Estimation for Massive MIMO-OTFS Systems}

	%
	%
	%
	
	\author{Kuo~Meng, Shaoshi~Yang, Xiao-Yang Wang, Yan~Bu, Yurong~Tang, Jianhua Zhang and Lajos~Hanzo
		\thanks{Copyright (c) 2015 IEEE. Personal use of this material is permitted. However, permission to use this material for any other purposes must be obtained from the IEEE by sending a request to pubs-permissions@ieee.org.}
\thanks{This work was supported in part by the Beijing Municipal Natural Science Foundation under Grant L202012 and Grant Z220004, in part by the BUPT-CMCC Joint Innovation Center under Grant S2021159, in part by the
Engineering and Physical Sciences Research Council projects
EP/W016605/1, EP/X01228X/1, EP/Y026721/1 and EP/W032635/1, and in part by the European Research Council's Advanced Fellow Grant QuantCom (Grant
No. 789028). \textit{(Corresponding author: Shaoshi Yang)}}		
		\thanks{K. Meng, S. Yang, and X.-Y. Wang are with the Key Laboratory of Universal Wireless Communications, Ministry of Education, Beijing University of Posts and Telecommunications, Beijing 100876, China (e-mail: \{meng\_kuo, shaoshi.yang, wangxy\_028\}@bupt.edu.cn).}
		\thanks{J. Zhang is with the State Key Laboratory of Networking and Switching Technology, Beijing University of Posts and Telecommunications, Beijing 100876, China (e-mail: jhzhang@bupt.edu.cn).}
		\thanks{Y. Bu and Y. Tang are with the Department
of Wireless and Terminal Technology, China Mobile Research Institute,
Beijing 100053, China (e-mail: \{buyan, tangyurong\}@chinamobile.com).}
		\thanks{L. Hanzo is with the School of Electronics and Computer Science, University of Southampton, UK (e-mail: lh@ecs.soton.ac.uk). }
	}

	\markboth{accepted to appear on IEEE Transactions on Vehicular Technology}%
	{Shell \MakeLowercase{\textit{et al.}}: Bare Demo of IEEEtran.cls for IEEE Journals}



	\maketitle
	
	\begin{abstract}
		We propose a channel estimation scheme based on joint sparsity pattern learning (JSPL) for massive multi-input multi-output (MIMO) orthogonal time-frequency-space (OTFS) modulation aided systems. By exploiting the \textcolor{black}{potential} joint sparsity of the delay-Doppler-angle (DDA) domain channel, the channel estimation problem is transformed into a sparse recovery problem. To solve it, we first apply the {{\textit{spike and slab}}} prior model to iteratively estimate the  support set of the channel matrix, and a higher-accuracy parameter update rule relying on the identified support set is introduced into the iteration. Then the specific values of the channel elements corresponding to the support set are estimated by the orthogonal matching pursuit (OMP) method.  Both our simulation results and analysis demonstrate that the proposed JSPL channel estimation scheme achieves an improved performance over the representative state-of-the-art baseline schemes, despite its reduced pilot overhead.
	\end{abstract}
	

	\IEEEpeerreviewmaketitle

	\section{Introduction}
	%
	%
	%
	%
	Massive multi-input multi-output (MIMO) schemes are expected to find their way into next-generation mobile systems, due to their unrivalled  advantages in spectral efficiency, energy efficiency and compact high-frequency form-factor \cite{7498075}. Massive MIMO aided orthogonal frequency division multiplexing (OFDM) exhibits excellent robustness against the frequency-selective fading\cite{Hu_2014}. However, in high mobility scenarios (e.g., high speed railway, unmanned aerial vehicles, etc.), the Doppler spread may cause severe inter-carrier interference (ICI), hence degrading the performance\cite{semiblind_2016}. 
	
	To overcome this challenge, the orthogonal time-frequency-space (OTFS) modulation was developed in \cite{7925924}, which outperforms OFDM in doubly-selective channels. The OTFS modulation scheme maps its transmitted symbols to the delay-Doppler (DD) domain rather than to the time-frequency (TF) domain of classic OFDM. For seamless integration into existing systems, OTFS can be implemented by adding a bespoke pre-processing module before the OFDM modulator, and a post-processing module at the receiver side. As a benefit, a time-varying channel can be converted into a near-time-invariant channel in the DD domain. Hence OTFS systems exhibit robust Doppler immunity in the face of high mobility and have attracted substantial research interests recently.
	
	An important issue for OTFS is the accuracy of channel estimation. For single-antenna OTFS systems, sophisticated impulse-based channel estimation methods were studied in \cite{8503182,8671740}. The author of \cite{8503182} suggested to use a pseudo-noise (PN) sequence for training, while Raviteja {\textit{et al.}} \cite{8671740} proposed an embedded pilot-aided channel estimation scheme requiring a guard region between pilot symbols. The performance of impulse-based methods are limited by the guard region and the pilot overhead, which makes it difficult to extend them to massive MIMO systems, where the pilot overhead is a major concern. To reduce the pilot overhead, the authors of \cite{9539066} and \cite{9456894} developed a joint pilot-and-data aided channel estimation algorithm. Srivastava {\textit{et al.}} \cite{Hanzo_otfs} proposed a guard-free pilot pattern and a sparse Bayesian learning (SBL) based channel estimation algorithm, and \cite{srivastava2021bayesian} explored the group sparsity for DD domains in MIMO-OTFS systems. Lei {\textit{et al.}} \cite{9184852} and \cite{zhao2022block} avoided guard region in SISO and small-scale MIMO systems with SBL methods using \textit{Laplace prior}. The existing studies on SBL have not investigated the  angular domain sparsity pattern, and the DDA domain joint sparsity in massive MIMO-OTFS systems also has to be researched. For massive MIMO-OTFS systems, Shen {\textit{et al.}} \cite{8727425} conceived a delay-Doppler-angle (DDA) domain sparse channel model and a 3D-structured orthogonal matching pursuit (SOMP) algorithm for downlink (DL) channel estimation. However, this scheme has a high pilot overhead, and assumes that the burst sparsity length of the angular domain is known in advance. Otherwise, the accuracy of the estimation is seriously affected. Liu {\textit{et al.}} \cite{9110823} exploited the reciprocity of time division duplex (TDD) and proposed an uplink-aided sparse DL channel estimation scheme, which reduces the pilot overhead. However, the application scenario of these contributions is limited, because they impose strong conditions, such as the \textit{a priori} knowledge of the sparsity pattern and the TDD-based reciprocity, for allowing their schemes to exploit the channel sparsity. 

	In this paper, we eliminate the above-mentioned strong assumptions and conceive an improved channel estimation algorithm based on joint sparsity pattern learning (JSPL) under the powerful Bayesian framework, for fully exploiting the potential sparsity of the DDA domain channel in massive MIMO-OTFS systems. For more accurately learning the sparsity pattern, we firstly apply a flexible {\textit{spike and slab}} prior model~\cite{6556987}, and derive its statistical channel model in the DDA domain under the Bayesian framework. Secondly, we propose a novel higher-accuracy  parameter update rule to capture the joint sparsity pattern in DDA domain for obtaining the support set of the channel matrix. Finally, we transform the channel estimation problem into a simpler sparse signal recovery problem thanks to the finite support set of the channel, and solve it by the OMP algorithm. Our analysis and simulation results demonstrate that the proposed algorithm achieves better performance than the representative state-of-the-art baseline schemes, despite its reduced pilot overhead.
	
	
	\section{System Model}
	
	\subsection{Time-Varying Massive MIMO Channel Model}
	We consider a single-cell massive MIMO-OTFS system having $N_\textrm{T}$ transmit antennas at the base station (BS) and $U$ single-antenna users. For DL transmission operating in the frequency division duplex (FDD) mode\footnote{In fact, the proposed scheme does not depend on  the specific duplex mode, hence it is also applicable to TDD.}, the users have to perform channel estimation and feed it back to the BS as a prerequisite for DL beamforming. Let us assume that the time-varying channel between a user and a BS antenna contains $N_{{\textrm{P}}}$ dominant propagation paths. Then the multi-path DL channel at the time instant $\kappa$ between the $(q+1)$th BS antenna ($q=0,1,2, \cdots,N_{\textrm T}-1$) and the $u$th user is given by
	\begin{equation}\label{eq1}
		h_{u,q,\kappa,\ell} =\sum \limits_{i = 1}^{{N_{{{\textrm P}}}}} {\alpha _{u,{i}}}{e^{j2\pi {\nu _{u,{i}}}\kappa {T_{{{\textrm s}}}}}}\delta \left( {\ell {T_{{{\textrm s}}}} - {\tau _{u,i}}} \right)e^{j 2\pi q{\psi _{u,i}}},
	\end{equation}
	where $\alpha _{u,i}$, $\nu _{u,{i}}$, $\tau _{u,i}$ and $\psi _{u,i}$ represent the propagation gain, Doppler \textcolor{black}{shift}, propagation delay and phase difference between two adjacent elements, corresponding to the $i$th path of the user $u$, respectively. Note that $u$, $\kappa$, $\ell$ and $q$ are all integer values, and $\ell$ represents the index of the propagation delay $\tau _{u,i}$ in terms of the number of $T_{\textrm s}$. We denote the angle of departure (AoD) corresponding to the $i$th path of the user $u$ as $\theta_{u,i}$. When a uniform linear array (ULA) of antennas is used, we have ${\psi_{u,i}} = \frac{d}{\lambda }\sin\theta_{u,i} $, where $d$ is the antenna spacing and $\lambda$ is the wavelength of the carrier frequency. Typically, $d = {\lambda/2} $ and $\theta_{u,i}\in[-\pi / 2, \pi / 2)$, thus we have ${\psi _{u,i}} \in [-1/2,1/2)$. Finally, $\delta(\cdot)$ denotes the Dirac delta function, and $T_{\textrm s}$ represents the sampling period or symbol duration.
	\subsection{OTFS Modulation and Demodulation}
	Consider a symbol sequence of length ${N_\ell}{N_k}$ that is rearranged into a two-dimensional data block $\mathbf{X}_q^{\rm DD}\in \mathbb{C}^{{N_\ell} \times {N_k}}$ at the $(q+1)$th antenna, where ${N_\ell}$ and ${N_k}$ represent the number of subcarriers and the number of symbols, respectively. The data block $\mathbf{X}^{\rm DD}_q$ supposed to be processed in the delay-Doppler (DD) domain is firstly converted to the time-frequency (TF) domain data block $\underline{\mathbf X}_q \in \mathbb{C}^{{N_\ell} \times {N_k}}$ by the inverse symplectic finite Fourier transform (ISFFT)\cite{7925924}. Then, $\underline{\mathbf X}_q $ is transformed into OFDM signal blocks by the inverse discrete Fourier transform (IDFT), followed by multiplying a transmitting window function $\mathbf{G}_{\textrm{TX}}\in\mathbb{C}^{{N_\ell}\times {N_\ell}}$. By inserting the cyclic prefix (CP) using the matrix $\mathbf{A}_{\textrm{CP}}\in\mathbb{C}^{({N_\ell}+N_{\textrm{CP}})\times {N_\ell}}$, the transmitted signal can be expressed as
	\begin{equation}	
		\mathbf{s}_{\it{q}}=\rm{vec} \{\mathbf{A}_{\textrm{CP}} \mathbf{G}_{\textrm{TX}}\mathbf{F}^{\rm{H}}_{\it{N}_{\ell}}\underbrace{\mathbf{F}_{\it{N}_{\ell}} \mathbf{X}^{\rm DD}_{\it{q}} \mathbf{F}^{\rm{H}}_{\it{N}_{k}}}_{ISFFT} \},
	\end{equation}where $N_{\textrm{CP}}$ is the length of CP, $\mathbf{F}_{{N_{\ell}}}\in\mathbb{C}^{{N_{\ell}} \times {N_{\ell}}}$ and $\mathbf{F}_{{N_k}}\in\mathbb{C}^{{N_k} \times {N_k}}$ are the DFT matrices, $(\cdot)^{\textrm H}$ denotes the conjugate transpose, and $\rm{vec}\{\cdot\}$ denotes the parallel-to-serial conversion.
	
	After the transmitted signal $\mathbf{s}_{\it{q}}$ passes through the channel, the signal $\mathbf{r}_{\it{u}}\in\mathbb{C}^{({N_\ell}+N_{\textrm{CP}}){N_k}\times 1}$ received  by the user $u$ is rearranged as a matrix $\mathbf{R}_u\in\mathbb{C}^{({N_\ell}+N_{\textrm{CP}})\times {N_k}}$, i.e., $\mathbf{R}_u =\rm{inverse\_{vec}} \{\mathbf{r}_{\it{u}} \}$. Then the signal removes the CP by using $\mathbf{A}_{\textrm{RCP}}\in\mathbb{C}^{{N_\ell}\times ({N_\ell}+N_{\textrm{CP}})}$ and passes through the receive window function $\mathbf{G}_{\textrm{RX}}\in\mathbb{C}^{{N_\ell}\times {N_\ell}}$. After applying the SFFT on the TF domain signal $\underline{\mathbf{Y}}_u \in \mathbb{C}^{{N_\ell}\times {N_k}}$, we obtain the DD domain data on the receiver side and it is expressed as
	\begin{equation}	
		\mathbf{Y}^{\rm DD}_u = {\mathbf{F}^{\rm{H}}_{\it{N}_{\ell}} \underbrace{\mathbf{F}_{\it{N}_{\ell}}\mathbf{G}_{\textrm{RX}} \mathbf{A}_{\textrm{RCP}} \mathbf{R}_{u}}_{\underline{\mathbf{Y}}_u}\mathbf{F}_{{N}_k}}.
	\end{equation}
	\subsection{Channel Input-Output Analysis} We denote the $(\ell+1,k+1+{N_k}/2)$th entry of  $\mathbf{Y}^{\rm DD}_u$ as ${y}^{\rm DD}_{u,\ell,k}$, which is expressed as \cite{8727425}
	\begin{align*}\label{eq7} 
		y^{\rm DD}_{u,\ell,k} \overset{N_k\rightarrow \infty }{=}
		&\sum _{q=0}^{N_{\textrm T}-1}\sum _{\ell ^{\prime }=0}^{{N_\ell}-1}\sum _{k^{\prime }=-N_k/2}^{N_k/2-1} 
		x^{\rm DD}_{\ell ^{\prime },k^{\prime },q}\\ &\times h^{\rm DDS}_{u,\ell -\ell ^{\prime },k-k^{\prime },q}e^{j2\pi \frac{\ell \left(k-k^{\prime }\right)}{N_k({N_\ell}+N_{\rm CP})}} +w^{\rm DD}_{u,\ell,k}.\tag{4} 
	\end{align*}
	\begin{figure}[t]\label{fig1}	\centering{\includegraphics[width=0.31\textwidth
			]{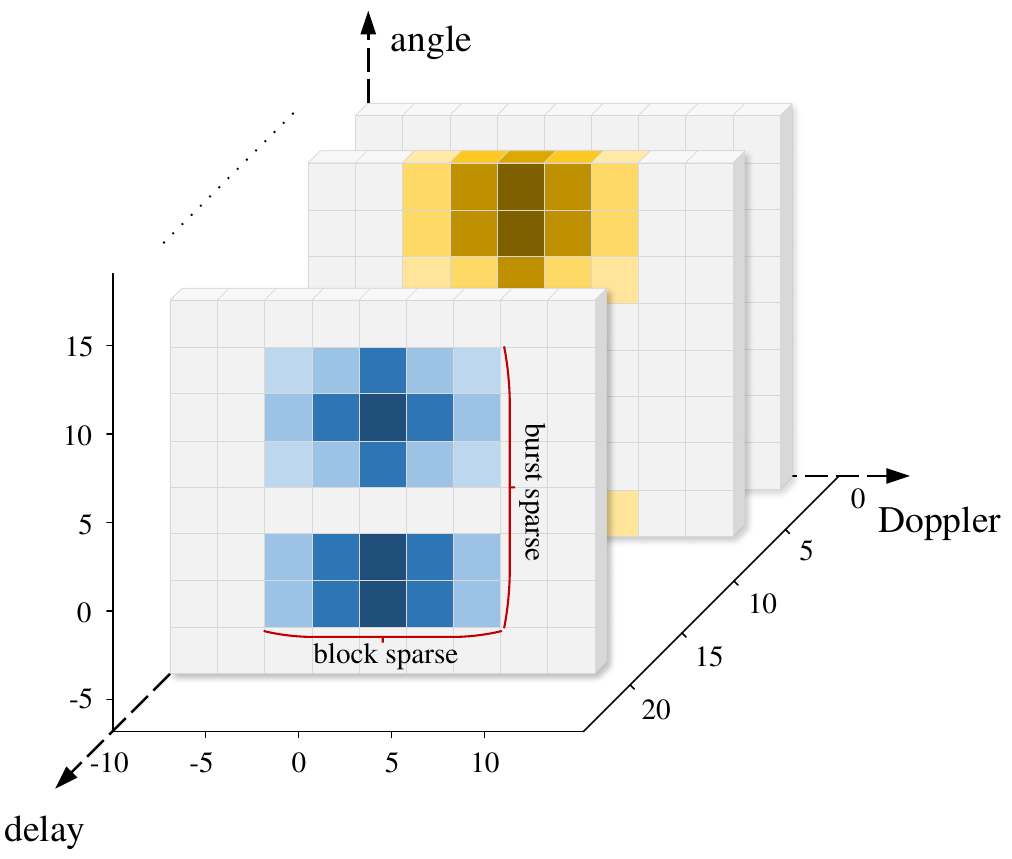}}	
		\caption{Sparsity pattern of the DDA domain channel model.}	
	\end{figure}Eq. \eqref{eq7} represents the two-dimensional convolution of the DD domain, $ h^{\rm DDS}_{u,\ell -\ell ^{\prime },k-k^{\prime },q}$ denotes the channel in the delay-Doppler-space (DDS) domain, and $w^{\rm DD}_{u,\ell,k}$ is the noise satisfying a complex Gaussian distribution with zero mean and variance $\eta$. From (\ref{eq1}) and (\ref{eq7}) and considering a \textit{spatial DFT} along the antenna index $q$ \cite{8187178}, we can obtain the DDA domain channel as
	\begin{align*}\label{dda_channel}
		h^{\rm DDA}_{u,\ell,k,r}
		=&\sum _{i=1}^{N_\textrm P}
		\alpha _{u,i}{e^{j2\pi {\nu _{u,{i}}} {T_{\textrm s}}}}\frac {\sin (\pi (\nu _{u,{i}}N_kT-k))}{\sin \left({\pi \frac {(\nu _{u,{i}}N_kT-k)}{N_k}}\right)} \\
		&\times e^{j\pi \frac {(\nu _{u,{i}}N_kT-k)(N_k-1)}{N_k}}\delta\left( {\ell {T_{{\textrm s}}} - {\tau _{u,i}}} \right)
		\\
		&\times \frac {\sin (\pi (N_\textrm T{\psi _{u,{i}}}-r))}{\sin \left({\pi \frac {(N_\textrm T{\psi _{u,{i}}}-r)}{N_\textrm T}}\right)}e^{j\pi \frac {(N_\textrm T{\psi _{u,{i}}}-r)(N_\textrm T-1)}{N_\textrm T}},\tag{5} 
	\end{align*}
	where $T=(N_\ell+N_{\textrm {CP}})T_\textrm s$, and $r\in \{-N_{\textrm T}/2,-N_{\textrm T}/2+1, \cdots, N_{\textrm T}/2-1\}$ denotes the index of AoD corresponding to the antenna index $q$. From (\ref{dda_channel}), we can verify that only few of the values of $h^{\rm DDA}_{u,\ell,k,r}$, are nonzero elements, which are obtained when $r \rightarrow N_\textrm T{\psi _{u,{i}}}$, $\ell = \tau_{u,i}N_{\ell}\Delta f$ and $k \rightarrow \nu _{u,{i}}N_kT$, where $\Delta f$ is subcarrier spacing. Thus the channel in the DDA domain is sparse.  Moreover, the channel is block-sparse in the Doppler domain and burst-sparse in the angle domain, as shown in {Fig. 1}. 
	\section{The Proposed JSPL Based Channel Estimation}
	\subsection{Derivation of JSPL Based Channel Estimation Algorithm}
	Rather than one-dimensional sparsity, e.g., angle domain sparsity \cite{7564429}, massive MIMO-OTFS channel often exhibits a joint sparsity pattern in multiple domains. The performance of channel estimation can be improved if this multi-dimensional sparsity is properly exploited. For the sake of simplicity, we arrange $h^{\rm DDA}_{u,\ell,k,r}$ into the column vector $\mathbf{h}_u\in\mathbb{C}^{N_\ell N_k N_{\textrm T}\times 1} $, where $\ell=0,1,\cdots,N_\ell-1,k=-N_k/2, -N_k/2+1, \cdots, N_k/2-1$, and $r=-N_{\textrm T}/2,-N_{\textrm T}/2+1, \cdots, N_{\textrm T}/2-1$. We may omit the subscript $u$ without confusion and convert Eq.~(\ref{eq7}) to 
	\begin{equation}\label{eq10}
		\mathbf{y}=(\mathbf{C}\star\mathbf{Z})\mathbf{h}+\mathbf{w}=\mathbf{\Phi}\mathbf{h}+\mathbf{w},\tag{6}
	\end{equation}
	where $(\mathbf{C}\star\mathbf{Z}) =\left[\mathbf{C}\odot \mathbf{Z}_{-\frac{N_{\textrm T}}{2}},  \mathbf{C}\odot\mathbf{Z}_{-\frac{N_{\textrm T}}{2}+1}, \cdots,   \mathbf{C}\odot\mathbf{Z}_r, \cdots 
	\notag\right.
	\\
	\left. 
	,\mathbf{C}\odot \mathbf{Z}_{\frac{N_{\textrm T}}{2}-1} \right]$, with $\odot$ denoting the Hadamard product. $\mathbf{Z}_r\in\mathbb{C}^{N_\ell N_k\times N_\ell N_k}$ is a two-dimensional convolution matrix, whose $[(\ell+1/2)N_k+k+1, (\ell^{\prime}+1/2) N_k+k^{\prime}+1]$th element equals $\sum_{q=0}^{N_{\textrm T}-1}e^{j2\pi\frac{qr}{N_{\textrm T}}}x^{\rm DD}_{\ell-\ell^{\prime },k-k^{\prime },q}$, 
	and $\mathbf{C}\in\mathbb{C}^{N_\ell N_k\times N_\ell N_k}$ is a matrix with the $\left[(\ell+1/2)N_k+k+1, (\ell^{\prime}+1/2) N_k+k^{\prime}+1\right]$th element being $ e^{j2\pi \frac{\ell \left(k-k^{\prime }\right)}{N_k(N_\ell+N_{{\textrm {CP}}})}}$. However, the sparsity pattern embedded in $\mathbf h$ is usually unknown \textit{a priori}. To exploit the sparsity pattern of $\mathbf h$, we invoke the flexible \textit{spike and slab} prior \cite{6556987} from the Bayesian perspective, thus expressing the \textit{a prior} distribution of $\mathbf h$ as 
	\begin{equation}\label{eq11}
		p({\mathbf{h}})
		=\prod_{n=1}^{N}\underbrace{[(1-\lambda_{n}) \delta(h_{n})+\lambda_{n} \mathcal {CN}(h_{n}\neq 0 ; 0,\mu)]}_{p(h_n)},\tag{7}
	\end{equation}where $\lambda_{n}\in(0,1)$ represents the probability of $h_{n}$ being nonzero\footnote{$\lambda_n$ essentially characterizes the sparsity degree of $\mathbf h$.}, $N=N_\ell N_kN_{\textrm T}=MN_{\textrm T}$, $h_n$ denotes the $n$th element of $\mathbf h$ and follows a complex Gaussian distribution ${\mathcal {CN}}(0, \mu)$ upon assuming Rayleigh fading except for $h_n = 0 $. The likelihood function of the received signal is given by
	\begin{equation}
		p(\mathbf{y}|\mathbf{h})={\mathcal{CN}}(\mathbf{y};\mathbf{\Phi} \mathbf{h}, \eta \mathbf{I}).\tag{8}
	\end{equation}
	Let ${\mathcal {N}}(x;\beta, \gamma)$ represent the expression $\frac{1}{ \sqrt{2 \pi\gamma}} e^{-\frac{(x-\beta)^{2}}{2 \gamma}}$. Then the posterior distribution of $\textrm{Re}\{h_{n}\}$ can be written as\footnote{Since our derivation aims to solve for the channel support set, namely the positions of the nonzero elements, rather than the value of the channel itself, it is sufficient to examine the real and imaginary parts of $h_n$ separately. Thus we can simplify the derivation and the execution of the JSPL algorithm by invoking the one-dimensional real Gaussian distribution $\mathcal {N}$ twice: one for the real part and the other for the imaginary part of $h_n$.} 
	\begin{equation}\label{eq13}
		p(\textrm{Re}\{h_{n}\}|\mathbf{y};\!\beta_n,\!\gamma_n)\!=\!\frac{1}{\zeta_{n}}p({\textrm{Re}\{h_{n}\}})\mathcal {N}(\textrm{Re}\{h_{n}\};\beta_n,\gamma_n)\tag{9}
	\end{equation}upon using the approximate message passing (AMP) algorithm \cite{amp},  which approximates
	the message passed from the node $h_n$ to all the nodes $y_m$ by a Gaussian distribution with a mean of $\beta_n$ and variance  of $\gamma_n$, according to the central limit theorem and factor graph. Here $y_m$, $m =1, \cdots, M$, is the $m$th element of $\mathbf{y}$, while $\beta_n$ and $\gamma_n$ are iteratively updated (see Algorithm 1), and $\zeta_{n}$ is the normalization factor defined as
	\begin{align*}\label{eq:zeta_n}
		\zeta_{n}
		&\triangleq\int_{\textrm{Re}\{h_{n}\}}p({\textrm{Re}\{h_{n}\}})\mathcal{N}(\textrm{Re}\{h_{n}\};\beta_n,\gamma_n)\\
		&=(1-\lambda_n)\mathcal{N}(0;\beta_n,\gamma_n)+\lambda_n \mathcal{N}(0;\beta_n,\gamma_n+\mu). \tag{10}
	\end{align*}
	Based on  (\ref{eq11}) and (\ref{eq:zeta_n}), we rewrite (\ref{eq13}) as
	\begin{small}
		\begin{align*}\label{eq11_true}
			p(\textrm{Re}\{h_{n}\}|\mathbf{y};\!\beta_n,\!\gamma_n)\!=&(1-\varphi_{n}) \delta(h_{n})\\
			&+\varphi_{n} \mathcal{N}\!\left(\!\textrm{Re}\{h_{n}\};\frac{\mu \beta_{n}}{\gamma_{n}+\mu} ,\frac{\mu \gamma_{n}}{\gamma_{n}+\mu} \right),\tag{11}
		\end{align*}
	\end{small}where we have
	  \begin{equation}\varphi_{n}=\frac{\lambda_{n}}{\zeta_{n}}\mathcal{N}(0;\beta_n,\gamma_n+\mu).\tag{12}
	  \end{equation}\label{eq:varphi_n}From \eqref{eq11_true} we can obtain the first and second order statistics of $\textrm{Re}\{h_{n}\}$, i.e., the mean and variance, as follows:
	\begin{align*}
		&\bar{h}_n=\varphi_{n}\frac{\mu \beta_{n}}{\gamma_{n}+\mu},\tag{13}\label{eq18}\\
		&v_n=\varphi_{n}\left[\left(\frac{\mu \beta_{n}}{\gamma_{n}+\mu}\right)^{2}+\frac{\mu \gamma_{n}}{\gamma_{n}+\mu}\right]-{\bar{h}_n}^{2}.\tag{14}\label{eq19}
	\end{align*}
	
	To exploit the statistical characteristics of the channel, we define the hyperparameters as $\boldsymbol{\rho}\triangleq\left\{ \boldsymbol{\lambda},\mu, \eta\right\}$, where $\boldsymbol{\lambda} = \{\lambda_n\}$. Let us denote the estimate of a parameter at the $t$th iteration by $(\cdot)^{(t)}$, $t\geq 1$. Then $\boldsymbol{\rho}^{(t)}$ can be obtained by maximizing the expectation of the joint probability density function $p({\mathbf h}, {\mathbf y})$ according to 
	\begin{align}\label{opt_problem}
		\boldsymbol{\rho}^{(t+1)} & =\arg \max _{\boldsymbol{\rho^{(t)}}} \mathrm{E}\left\{\ln p(\mathbf{h}, \mathbf{y}) | \mathbf{y} ; \boldsymbol{\rho}^{(t)}\right\} \nonumber \\ 
		& = \arg \max _{\boldsymbol{\rho^{(t)}}} \int_{\mathbf{h}}\ln [p({\mathbf{h}})p(\mathbf{y}|\mathbf{h})]p(\textrm{Re}\{\mathbf{h}\}|\mathbf{y};\boldsymbol{\rho}^{(t)}), \tag{15}
	\end{align}
	where $p(\textrm{Re}\{\mathbf{h}\}|\mathbf{y};\boldsymbol{\rho}^{(t)})=\prod_{n}p(\textrm{Re}\{h_{n}\}|\mathbf{y};\beta_n,\gamma_n)$. Since jointly optimizing all the variables in $\boldsymbol{\rho}$ is complex, we use the incremental expectation-maximization (EM) rule of\cite{6556987}  to update one variable at a time,  while fixing the others. Firstly, to capture the channel's sparsity pattern, we update $\{\lambda_{n}\}$ for the above optimization problem as follows:
	\begin{align}\label{eqno16}
		\lambda_{n}^{(t+1)} & =\arg\max_{\lambda_n \in(0,1)}  \mathrm{E}\left\{\ln [p(\mathbf{h}) p(\mathbf{y}| \mathbf{h})] | \mathbf{y} ; \boldsymbol{\rho}^{(t)}\right\} \nonumber \\
		& =\arg\max_{\lambda_n \in(0,1)} \int_{\mathbf{h}}[\ln p(\mathbf{h})+C] p(\operatorname{Re}\{\mathbf{h}\} | \mathbf{y} ; \boldsymbol{\rho}^{(t)}) \nonumber\\
		& =\arg\max_{\lambda_n \in(0,1)} \int_{\mathbf{h}} \ln p(\mathbf{h}) p(\operatorname{Re}\{\mathbf{h}\} | \mathbf{y} ; \boldsymbol{\rho}^{(t)}), \tag{16}
	\end{align} where $C$ denotes a constant, and \eqref{eqno16} implies that for the update of $\lambda_{n}$, $p(\mathbf{y}| \mathbf{h})$ can be considered as a constant term to be ignored. To solve the problem \eqref{eqno16}, we obtain
\begin{align}\label{eqno17}
\frac{d}{d \boldsymbol{\lambda}} \int_{\mathbf{h}} \ln p(\mathbf{h}) p\left(\operatorname{Re}\{\mathbf{h}\} | \mathbf{y} ; \boldsymbol{\rho}^{(t)}\right)=0. \tag{17}
\end{align}
By means of Leibniz's formula and some approximate operations, we convert $\eqref{eqno17}$ to:
\begin{align}\label{eqno18}
	\sum_{n=1}^{N} \int_{h_{n}} p\left(\operatorname{Re}\left\{h_{n}\right\} \mid \mathbf{y} ; \beta_{n}, \gamma_{n}\right) \frac{d}{d \lambda_{n}} \ln p(h_{n})=0, \tag{18}
\end{align}
where
\begin{align}\label{eqno19}
	\frac{d}{d \lambda_{n}} \ln p\left(h_{n}\right) & = \frac{\mathcal{C N}\left(h_{n} \neq 0 ; 0, \mu\right)-\delta\left(h_{n}\right)}{p\left(h_{n}\right)} \nonumber\\
	& =\left\{
		\begin{array}{ll}
		\frac{1}{\lambda_{n}}, & h_{n} \neq 0. \\
		-\frac{1}{1-\lambda_{n}}, & h_{n}= 0.
		\end{array}
	\right.\tag{19}
\end{align}
To solve the integral term in $\eqref{eqno18}$, we define an interval $\mathcal{Z} \triangleq[-\varepsilon, \varepsilon]$, and represent the complement of $\mathcal{Z}$ by $\overline{\mathcal{Z}}$ in the real field. When $\varepsilon \rightarrow 0$ we have: 
\begin{align}
\lim _{\varepsilon \rightarrow 0} \int_{\operatorname{Re}\left\{h_{n}\right\} \in \mathcal{Z}} p\left(\operatorname{Re}\left\{h_{n}\right\} | \mathbf{y} ; \beta_{n}, \gamma_{n}\right)&=1-\varphi_{n},\tag{20}\label{eqno20}\\
\lim _{\varepsilon \rightarrow 0} \int_{\operatorname{Re}\left\{h_{n}\right\} \in \overline{\mathcal{Z}}} p\left(\operatorname{Re}\left\{h_{n}\right\} | \mathbf{y} ; \beta_{n}, \gamma_{n}\right)&=\varphi_{n}.\tag{21}\label{eqno21}
\end{align}
By combining \eqref{eqno19}, \eqref{eqno20} and \eqref{eqno21}, we can derive \eqref{eqno18} as: 
\begin{align}
\sum_{n=1}^{N}\left[\frac{1}{\lambda_{n}} \varphi_{n}-\frac{1}{1-\lambda_{n}}\left(1-\varphi_{n}\right)\right]=0. \tag{22}
\end{align}
Hence, the update formula for $\lambda_{n}$ is obtained as:
\begin{align}\label{eqno23}
	\lambda_{n}^{(t+1)}=\varphi_{n}^{(t)}.\tag{23}
\end{align}From \eqref{eqno23} we find that the update of $\{\lambda_{n}\}$ only depends on the $n$th element of the channel vector $\mathbf{h}$, which means it does not capture the joint sparsity pattern of the DDA channel.
\begin{algorithm}[t]
	\footnotesize
	\renewcommand{\algorithmicrequire}{\textbf{Input:}}
	\renewcommand{\algorithmicensure}{\textbf{Output:}}
	\caption{\footnotesize{JSPL-Based Channel Estimation Algorithm}}
	
	\label{alg:1}
	\begin{algorithmic}[1]
		
		\REQUIRE $\mathbf{y}$, $\mathbf{\Phi} = \{{\it{\Phi}}_{m n}\}$
		\renewcommand{\algorithmicrequire}{\textbf{Initialization:}}
		\REQUIRE $t=1$, $i=1$, $\Omega=\emptyset$, $\Omega_\textrm{d}=\emptyset$, $\Omega_\textrm{DA}=\emptyset$,  $S_{m}^{(0)} =0$, $N_{\textrm{P}} = 0$, $T_{\textrm {MAX}}$, $\boldsymbol{\rho}^{(1)}$, $\epsilon_1$, $\epsilon_2$, $\bar{h}_{n}^{(1)}$, $v_n^{(1)}$

		\STATE Update process parameters:\\\label{k1}
		$\quad V_{m}^{(t)}=\sum_{n}\left|{\it{\Phi}}_{m n}\right|^{2} v_n^{(t)}$\\ 
		$\quad S_{m}^{(t)}=(y_m-\sum_{n} {\it{\Phi}}_{m n} \bar{h}_{n}^{(t)}+{V_{m}^{(t)}}S_{m}^{(t-1)})/({\eta^{(t)}+V_{m}^{(t)}})$
		\STATE Update variable parameters:\\
		$\quad \gamma_{n}^{(t)}=\left[\sum_{m} \frac{\left|\it{\Phi}_{m n}\right|^{2}}{\eta^{(t)}+V_{m}^{(t)}}\right]^{-1}$ \\
		$\quad \beta_{n}^{(t)}=\bar{h}_{n}^{(t)}+\gamma_{n}^{(t)} \sum_{m} {\it{\Phi}}_{mn}S_{m}^{(t)}$
		
		\STATE Update $\varphi_{n}^{(t)}$, $\bar{h}_{n}^{(t+1)}$ and $v_n^{(t+1)}$ with (12), \eqref{eq18} and \eqref{eq19} 

		\STATE Update $\boldsymbol{\rho}^{(t+1)}$ with \eqref{eq22}, \eqref{eq23} and \eqref{eq24}
		\STATE Let $t=t+1$ and proceed to Step \ref{k1} until $\left\|\boldsymbol{\lambda}^{(t+1)}-\boldsymbol{\lambda}^{(t)}\right\|_{2}<\epsilon_1\left\|\boldsymbol{\lambda}^{(t)}\right\|_{2}$ or $t=T_{\textrm {MAX}}$, s.t. $\forall n\in[1,\cdots, N]$ 
		\STATE Reshape $\boldsymbol{\lambda}$ as a tensor $ \mathbf{\Lambda}$ 
		\IF{$ \left\|\mathbf{\Lambda}(\ell,:)\right\|_2>\epsilon_2$}
		\STATE  
		$N_{\textrm P}=N_{\textrm P}+1$
		\STATE  $\Omega_{\textrm{d}}=\Omega_{\textrm{d}}\cup\ell$
		\ENDIF
		\FOR{$i\leq N_{\textrm P}$}
		\IF{$ \mathbf{\Lambda}(\ell_\textrm{d}^{(i)},k,r)>\epsilon_2$, s.t. $\ell_\textrm{d}^{(i)}\in\Omega_{\textrm{d}}$}
		\STATE  $\Omega_{\textrm{DA}}^{(i)}=\Omega_{\textrm{DA}}^{(i)}\cup(k,r)$
		\ENDIF
		\STATE $\Omega=\Omega\cup(\ell_\textrm{d}^{(i)},k,r)$, s.t. $\forall (k,r)\in\Omega_{\textrm{DA}}^{(i)}$
		\STATE $\mathbf{h}^{(i)}|_{\underline{\Omega}}={\mathbf{\Phi}}^{\dagger}|_{\underline{\Omega}} \times \mathbf{y}$, where $(\cdot)^\dagger$ denotes pseudo-inverse and $\underline{\Omega}$ indicates the positions of channel supports determined by $\Omega$. 
		\STATE $i=i+1$
		\ENDFOR
		\ENSURE $\mathbf{h}$
	\end{algorithmic}  
\end{algorithm} 
By considering the multi-burst sparsity of the angle domain, the supports of the channel exhibit two-dimensional multi-block sparsity in the Doppler-angle domain, which implies that the Doppler domain channel supports and the angular domain channel supports intersect across multiple blocks in the Doppler-angle plane. Therefore, we update  $\{\lambda_{n}\}$ by using the  adjacency-assisted method, as detailed below, in order to learn the sparsity pattern of the channel, and propose the following update rule:
	\begin{align*}\label{eq22}
		\lambda_{n}^{(t+1)}=\frac{1}{\sum_{a}\xi_{n,a}|\Gamma(n,a)|} \sum_a \sum_{b=1}^{|\Gamma(n,a)|}
		 \xi_{n,a}\varphi_{n,a,b}^{(t)}, \tag{24}
	\end{align*}where $\Gamma(n,a)$ denotes the set of indices of the $a$th nearest elements in $\mathbf{h}$ for $h_{n}$. For example, the coordinate of $h_{n}$ in the Doppler-angle domain is $(k,r)$, then the set of its nearest elements ($a=1$) is denoted by $\Gamma(n, 1)=\{(k-1, r),(k+1, r),(k,(r-1+N_{\mathrm{T}})_{N_{\mathrm{T}}}),(k,(r+1+N_{\mathrm{T}})_{N_{\mathrm{T}}})\}$ as shown in Fig. 2, and ${( \cdot )_{{N_{\rm{T}}}}}$ represents modulo $N_{\rm{T}}$. Case 1 represents the most conventional set of nearest neighbors containing the left and right elements of the current element in the Doppler domain and the up and down elements in the angle domain. Case 2 represents the set of nearest neighbors of a max Doppler boundary element containing the left element of the current element in the Doppler domain and the up and down elements in the angle domain. While Case 3 represents the set of nearest neighbors of an angle domain boundary element containing the left and right elements of the current element in the Doppler domain and the lower element and the cyclic mode-taking element due to the burst sparsity pattern in the angle domain.
$\varphi_{n,a,b}^{(t)}$ represents the particular $\varphi_n^{(t)}$ corresponding to the $b$th element of $\Gamma(n,a)$, $\xi_{n,a}$ denotes the weight of $\varphi_{n,a,b}^{(t)}$, and $|\Gamma(n,a)|$ is the cardinality of $\Gamma(n,a)$.
	\begin{figure}[t]
	\centering{	\includegraphics[
		width=0.31\textwidth]{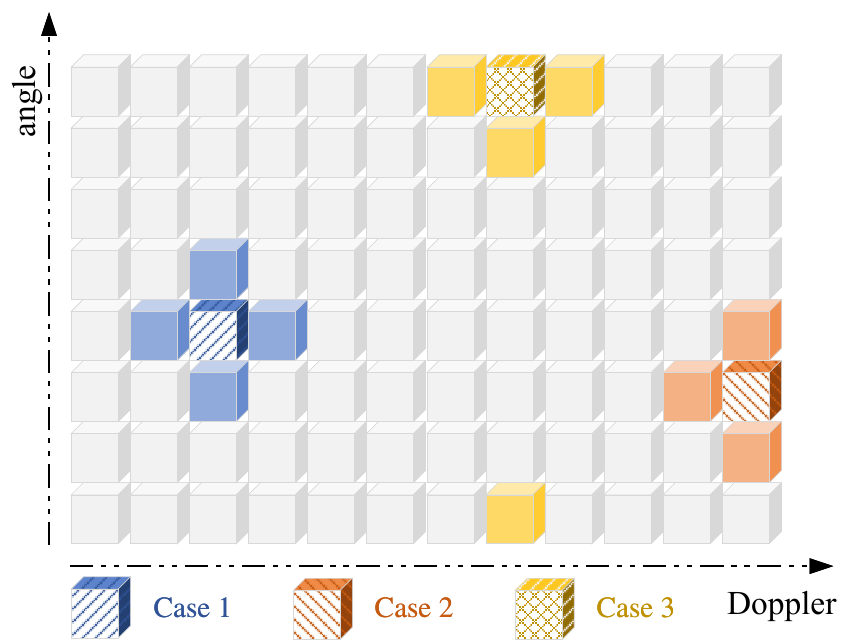}}
\caption{Three cases of the set of the nearest elements.}
\end{figure} 
Secondly, the derivative of the objective function of \eqref{opt_problem} is calculated with respect to the individual parameters $\mu$ and $\eta$, which are then updated relying on the EM \cite{6556987} as
\begin{small}
		\begin{align*}
			\mu ^{(t+1)}&=\frac{1}{\sum_{n} \varphi_{n}^{(t)}}\!\sum_{n=1}^N\! \varphi_{n}^{(t)}\!\left[\left(\frac{\mu^{(t)} \beta_{n}^{(t)}}{\gamma_{n}^{(t)}+\mu^{(t)}}\right)^{2}\!\!+\frac{\mu^{(t)} \gamma_{n}^{(t)}}{\gamma_{n}^{(t)}+\mu^{(t)}}\right], \tag{25}\label{eq23}\\
			\eta^{(t+1)}&=\frac{1}{M} \sum_{m=1}^M\left[\left({\eta^{(t)}S_{m}^{(t)}}\right)^2+\frac{\eta^{(t)} V_{m}^{(t)}}{\eta^{(t)}+V_{m}^{(t)}}\right],\label{eq24} \tag{26}
		\end{align*}
\end{small}where $V_{m}^{(t)}$ and $S_{m}^{(t)}$ are intermediate parameters updated first in each iteration of our algorithm. Then, $\gamma_{n}^{(t)}$, $\beta_{n}^{(t)}$, $\varphi_{n}^{(t)}$, $\bar{h}_{n}^{(t+1)}$ and $v_n^{(t+1)}$ are updated in turn. Next, we update $\boldsymbol{\rho}^{(t+1)}$ to capture the sparsity pattern of the channel by setting an appropriate threshold $\epsilon_1$ for ${\boldsymbol{\lambda}}$. Thus we can convert the channel estimation problem into a sparse signal recovery problem based on the DDA domain channel supports. Specifically, we first obtain the channel supports in the delay domain, then by traversing each path we filter out the Doppler-angle domain two-dimensional multi-block channel supports corresponding to the path, and finally obtain the estimated channel. For clarity, the above procedures are summarized in Alg. \ref{alg:1}.

	\subsection{Analysis of Computational Complexity and Pilot Overhead}
	The computational complexity of our JSPL algorithm is mainly determined by two parts. Firstly, the complexity of the hyperparameter learning part is determined by the update of parameters, which is characterized by $\mathcal{O}(T_{\textrm {MAX}}MN)$, where $T_{\textrm {MAX}}$ denotes the maximum number of iterations and it is much smaller than $MN$. Secondly, in the sparse signal recovery part, the complexity is dominated by calculating the pseudo-inverse of $\mathbf{\Phi}$. According to compressed sensing theory~\cite{1614066}, the actual dimension of the matrix involved in the inverse operation is $M \times Q$, where $Q$ indicates the sparsity of the DDA domain channel and it is given by $Q=N_{\textrm P}D_{{\textrm {MAX}}}N_\textrm{D}$, with $D_{{\textrm {MAX}}}$ being the index corresponding to the maximum Doppler \textcolor{black}{shift} and $N_{\textrm{D}}\approx N_{\textrm T}/10$ being the length of a single burst in the angle domain~\cite{7564429}. Thus when $M>Q$, the matrix has a left inverse, at which point the complexity is $\propto$ max$\{Q^3,Q^2M\}=\mathcal{O}({Q^2M})$. While when $M<Q$, the complexity is $\mathcal{O}({QM^2})$, so the complexity of the inverse calculation is $\propto$ max$\{\mathcal{O}(Q^2M),\mathcal{O}(QM^2)\}$. Considering the increased $N_{\textrm T}$ in massive MIMO systems, $Q$ represents the dominant contribution. Thus the complexity of JSPL is $\mathcal{O}(\frac{N_{\textrm P}D_{{\textrm {MAX}}}}{10}{N_\ell}^2{N_k}^2N_{\textrm T}+T_{\textrm {MAX}}{N_\ell}^2{N_k}^2N_{\textrm T})$, where $N_{\textrm P}$ is typically much smaller than the other parameters. In massive MIMO systems, the complexity of JSPL is linearly increasing with $N_T$, thus it can be controlled within acceptable limit by right-sizing the OTFS resource blocks.
	
	For the impulse-based channel estimation scheme,  \cite{8671740}, the pilot overhead is $\propto {N_\ell}{N_k}N_{\textrm T}$, and the guard regions occupy substantial spectral resources. Additionally, upon increasing of the pilot overhead, the guard regions between different impulses may become insufficient, thus resulting in mutual interference. Therefore, it is unsuitable for massive MIMO systems. By contrast, the pilot overhead of channel estimation schemes based on sparse signal recovery is typically $\propto Q\log({N_\ell}{N_k}N_{\textrm T})$, which is a lower bound. However, in reality the number of  dominant path $N_{\textrm P}$ and the multi-block sparsity of the Doppler-angle domain are usually unknown. As a result, the method of determining channel supports by correlation\cite{8727425} often requires using more pilots than the lower-bound value to ensure high accuracy. Notably, our proposed JSPL scheme accurately estimates the channel supports by means of parameter learning, and the learning process can be offline nature, hence an ultra-low pilot overhead is incurred.

	\section{Simulation Results and Discussions}
	In this section, we compare the performance of our proposed JSPL algorithm and representative baseline schemes by numerical simulations. We use the 3GPP standardized channel model containing 6 dominant paths\cite{3gpp}. The carrier frequency is 4.9 GHz, the Duplex mode is FDD and the subcarrier spacing is $\Delta f=15\text{kHz}$ . The number of BS antennas is $N_{\textrm T}=64$, and the UE has a single antenna. Then We set the size of the OTFS resource block to $(N_{\ell},N_k)=(1024, 128)$, and $\text{CP}=16.6 \text{us}$. Other initializations concerning the learning can be found in Alg.~\ref{alg:1}. Finally, the receiver uses the popular message passing (MP) based algorithm relying on threshold detection~\cite{8671740}. 
	
	In Fig. 3, we compare the JSPL recovery (\textcolor{black}{on} the right) \textcolor{black}{with} the true channel (\textcolor{black}{on} the left) as a function of the degree of sparsity  $\lambda_{n}$ -- indicated by the \textcolor{black}{accompanying} color bar -- of the Doppler-angle domain channel matrix. The channel supports are obtained by comparing the JSPL recovery to the threshold $\epsilon_2$ (see Step 12 of Alg. \ref{alg:1}). We observe that the accuracy of the two-dimensional channel support recovery in the Doppler-angle domain is high.
	
	In Fig. 4, we evaluate the normalized mean square error (NMSE) performance of JSPL against the baseline schemes. Two vehicular speeds are considered: 120km/h and 360km/h. Since the impulse-based scheme is constrained by the guard region, while OMP and 3D-SOMP lack \textit{a priori} knowledge about the number of dominant propagation paths and the channel supports, they are all inferior to our JSPL. Additionally, the performance improvement of JSPL at  120km/h is higher than at 360km/h. This is because the capability to estimate the Doppler spread for a given number of OTFS resource blocks tends to be degraded for a higher velocity.
	\begin{figure}[t]\label{fig2}
		\centering{	\includegraphics[
			width=0.31\textwidth]{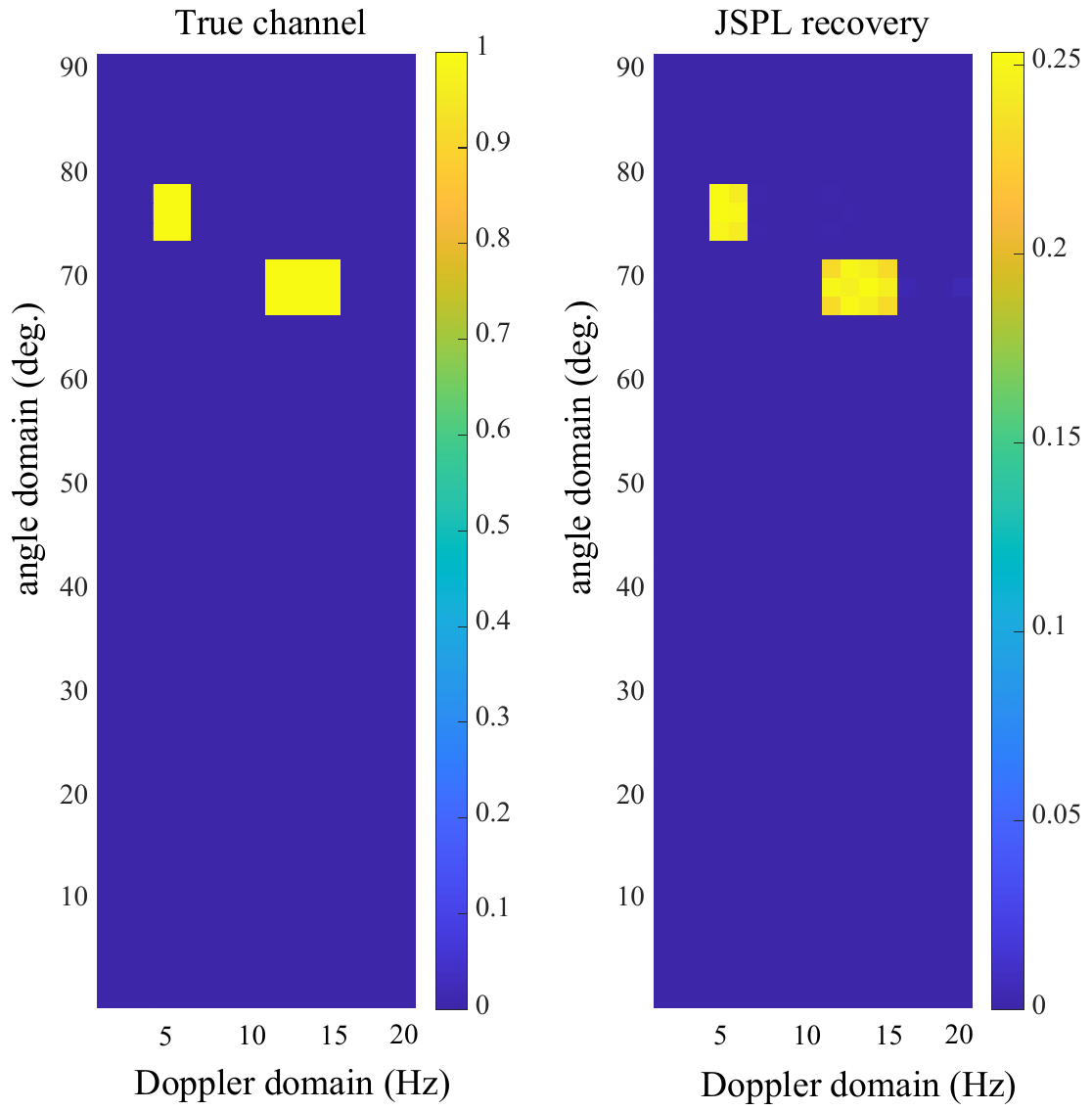}}
	\caption{\textcolor{black}{The JSPL recovery (on the right) and the true channel (on the left)} vs. the sparsity degree $\lambda_{n}$ of the Doppler-angle domain channel matrix.}
\end{figure} 

Fig. 5 compares the NMSE performance of our JSPL and the baseline schemes under different pilot overheads and signal-to-noise ratio (SNR) values. We set the user's speed to 360km/h. It is observed that although our JSPL uses only 10\% of the overhead of the baseline algorithms, it still outperforms the baselines thanks to its ability to learn the channel supports. In other words, our JSPL substantially reduces the pilot overhead, while ensuring a higher estimation accuracy than the baseline sparse signal recovery algorithms relying on correlation-aided channel support judgement. Moreover, the performance improvement of our JSPL becomes small with high pilot overhead, because the accuracy of determining channel supports by correlation is also significantly improved when there are sufficient pilots. However, such a high pilot overhead is unaffordable to massive MIMO systems.

Finally, in Fig. 6 we compare the bit error rate (BER) upon using different channel estimation algorithms and perfect channel knowledge. We set the user's speed to 360km/h and the pilot overhead to 50\% for the baseline channel estimation algorithms (5\% for JSPL and 0 for the perfect channel). It can be seen that our JSPL still outperforms the baseline channel estimation algorithms and exhibits a small gap w.r.t. the lower-bound provided by the perfect channel assumption.

\begin{figure}[t]
	\centering{	\includegraphics[width=0.38\textwidth]{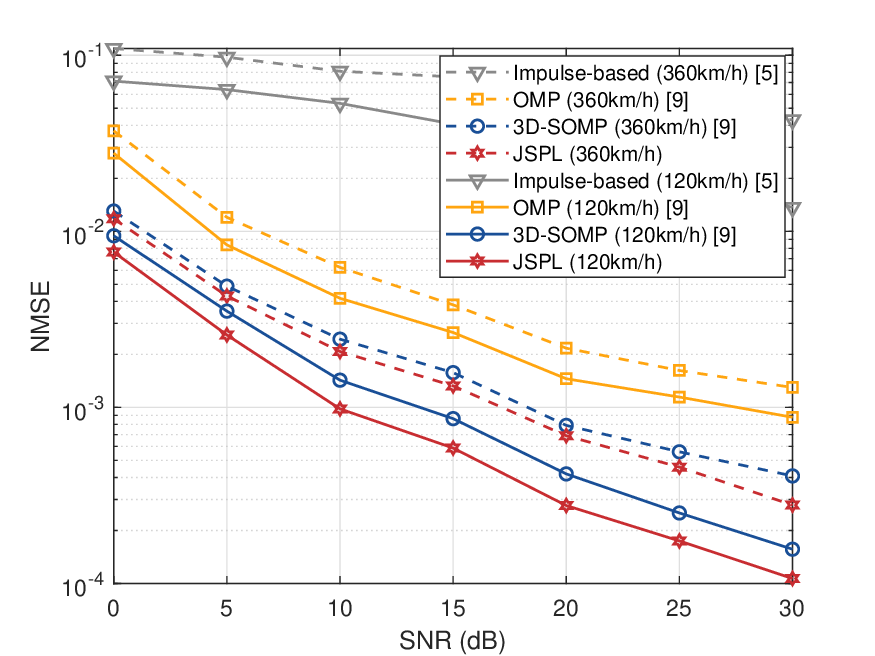}}
\caption{NMSE of our JSPL and the baseline schemes at different speeds.}
\end{figure} 
\begin{figure}[t]
\centering{	\includegraphics[width=0.38\textwidth]{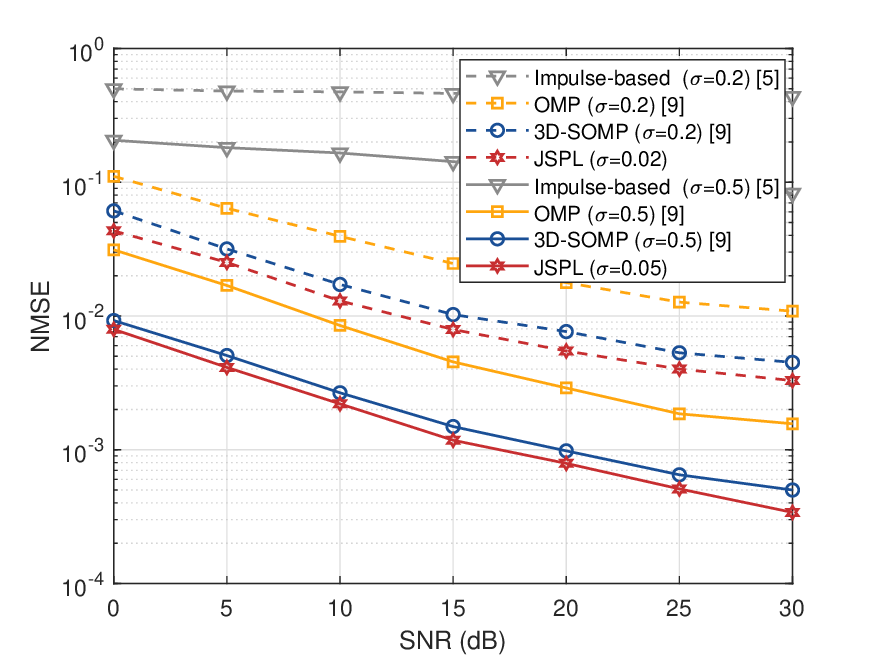}}
\caption{NMSE of our JSPL and the baseline schemes with different pilot overhead $\sigma = \frac{\textrm{Number of resource blocks occupied by pilots}}{MN} $.}
\end{figure} 
\begin{figure}[t]
\centering{	\includegraphics[width=0.38\textwidth]{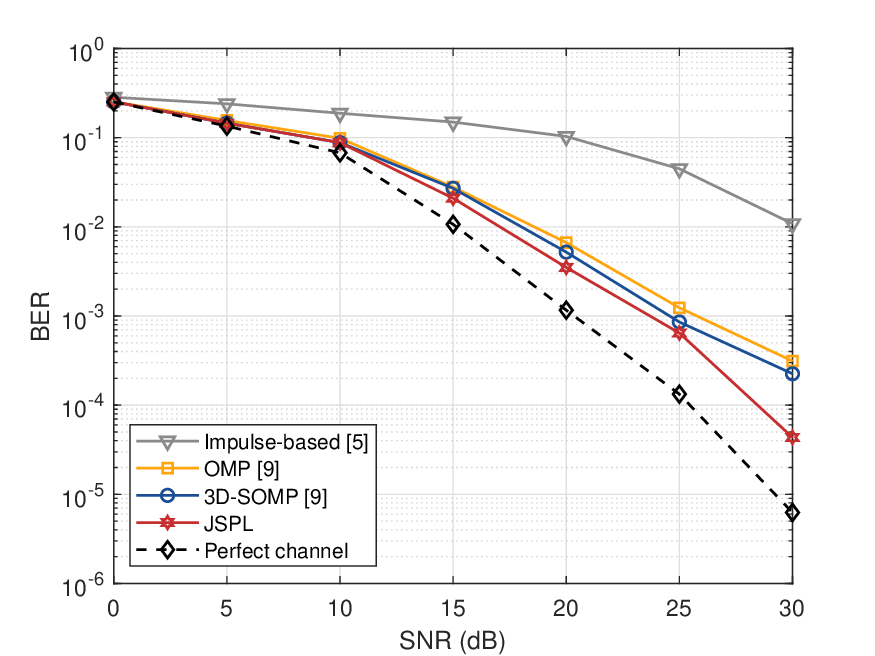}}
\caption{BER of our JSPL and the baseline schemes.}
\end{figure} 
\section{Conclusion}
We have proposed a JSPL based channel estimation algorithm for massive MIMO-OTFS systems under a Bayesian framework by exploiting the \textcolor{black}{potential} joint sparsity of the DDA domain channel. Due to the \textcolor{black}{uncertainty of sparsity in realistic} channels, we apply the \textit{spike and slab} prior model to fit the channel and propose a new parameter update rule to estimate the channel \textcolor{black}{support set} by exploiting the channel's joint sparsity \textcolor{black}{in multiple domains}. Then we solve a sparse signal recovery problem based on the support set of the channel. Our analysis and simulation results demonstrate that the proposed algorithm exhibits competitive performance and greatly reduced pilot overhead at the cost of moderately increased computational complexity.


%

\ifCLASSOPTIONcaptionsoff
\newpage
\fi



\bibliographystyle{IEEEtran}
\bibliography{cite.bib}
%

\end{document}